# A dual tunnel structure for the Einstein Telescope

Riccardo DeSalvo[1], Leonardo Gonzalez[2], Nelson Leon[1], Marina Mondin[1], Harry Themann[1], Fabian Peña Arellano[1]

1 California State University, Los Angeles, Department of Engineering, 5151 State University Drive, Los Angeles CA 90032

2 University of Guadalajara, University Centre for Exact Sciences and Engineering, (CUCEI), Blvd. Gral. Marcelino Garcıa Barragan 1421, Col. Olımpica, 44430, Guadalajara, Jalisco, Mexico.

## Abstract

We present a novel tunnel architecture for the Einstein Telescope that departs from the traditional large-cavern approach and reduces the excavated volume by an order of magnitude. In the proposed design, seismic isolation systems are housed in raise-bore wells extending upward from the main tunnel toward an upper service tunnel. The pre-isolators for the most sensitive optics are located in the service tunnel, supported directly on strong rock**,** whereas the other filters are distributed along the wells within compact, side-access vacuum chambers. Shorter, separate wells accommodate the seismic isolation systems for less demanding optics. This configuration provides substantial advantages: easier lock acquisition and improved robustness of the interferometers, lower-frequency pendulum stages, reduced congestion around the test masses, simplified installation and maintenance, improved vacuum partitioning, strong physical decoupling between the high- and low-frequency interferometers, and compatibility with advanced sensors for Newtonian-noise cancellation.

## Introduction

The concept of the Einstein Telescope (ET), a third-generation underground gravitational-wave observatory, was first proposed approximately 25 years ago, when Virgo was still in its early operational phase[1]. The initial design consisted of an equilateral triangle, with 10-km-long sides and three interferometric detectors. Each detector was subsequently conceived as a xylophone, comprising a high-power, room-temperature instrument optimized for high-frequency operation (ET-HF) and a cryogenic instrument (ET-LF) extending sensitivity toward the lowest frequencies [2]. The former must contend primarily with the challenges associated with the large circulating optical power needed to minimize shot noise, while the latter is fundamentally limited at low frequencies by effects such as Newtonian noise, radiation-pressure noise, mirror and suspension thermal noise, and residual seismic noise. Newtonian noise arises from fluctuations in Earth's gravitational field produced by density perturbations associated with seismic waves and can therefore mimic the effects of gravitational waves themselves[3]. Radiation-pressure noise reflects statistical fluctuations in the intracavity photon number. Suspension thermal noise originates from thermal fluctuations associated with mechanical losses at the points of flexure[4]. Coating thermal noise results from the statistical fluctuations of elastic energy in the mirror substrate and its reflective coatings[5].

More recent studies have considered replacing the 10-km triangular configuration with two L-shaped detectors, each with 15-km-long arms, separated by at least 1000 km and rotated by approximately 45° relative to one another to ensure full coverage of both gravitational-wave polarizations.[6]

The main motivations for placing ET underground, at a depth of roughly 300 m, are reduced Newtonian noise and improved environmental stability. Optimal performance for Newtonian Noise cancellation requires locating the test masses within strong, homogeneous rock and as far as practically possible from the boundary between hard rock and the softer alluvial overburden.

This work presents an alternative tunnel-and-well configuration for the end and beam splitter stations intended to improve observatory performance while substantially reducing excavation volume.

## The origin of the large cavern concept

Historically, the ET was envisioned as a scaled-up, underground version of Virgo[7], with large caverns housing the seismic isolation towers inside tall vacuum chambers. The size of those caverns was determined by the considerable height required for the isolation chains that isolate the test masses. At the time, limited operational experience with large underground interferometers naturally favored an architecture closely derived from the Virgo concept, while alternative configurations have received less consideration.

The tunnel design adopted by KAGRA,[8] illustrated in Figure 1, marked the first significant departure from this paradigm. A service tunnel was placed above the main interferometer tunnel to relocate the anchoring points of the seismic isolation chains away from the interferometer plane. This approach offered several advantages: decluttering the volume surrounding the test masses, enabling longer isolation chains, and improving the stability of the upper stage of the seismic isolation chain by seating it directly on hard rock, without the amplification effects found in all tall structures. However, KAGRA still relied on large caverns to house the large, side-access, walk-in cryostat, limiting the extent to which the surrounding rock could be exploited for mechanical support and isolation.

More recent cost-reduction studies have considered reducing the size of the ET caverns by approximately 25% [9]. Such a reduction necessarily limits the available height for the seismic isolation systems, potentially reducing the margin available in residual seismic and suspension thermal noise. Such a design may also constrain the feasibility of future upgrades, including improved Newtonian-noise mitigation techniques.

## The dual tunnel concept

We propose an alternative architecture for the end and beam splitter stations designed to improve the performance of both the low frequency and high frequency detectors while providing stronger physical separation between the anchoring points of the seismic isolation chains, leaving only the optical components in the main tunnel. This separation is also designed to enable maintenance or modifications to one interferometer without interfering with the astrophysical observations in the other. Local enlargement of the main tunnel may still be required, for example to meet Newtonian-noise requirements around the ET-LF test masses, if studies indicate that a cylindrically symmetric cavity around them would be advantageous.

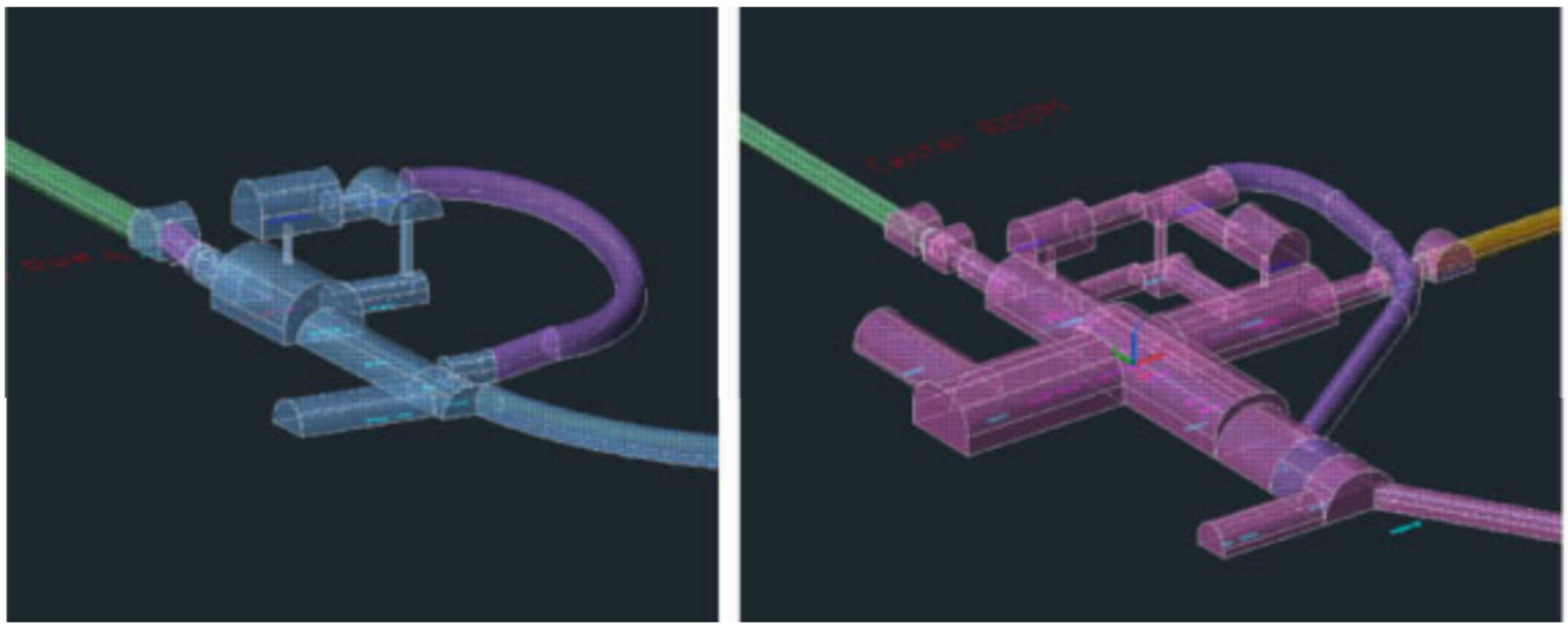

*Figure 1: Dual tunnel configuration of one end station (left) and of the beam splitter station (right) in KAGRA. The heads of the test-mass seismic isolation chains are seated on the rock floor of the upper tunnel, 12 m above the interferometer plane. The seismic isolation systems for all other optics are still housed within large underground caverns.*

The proposed structure is illustrated in Figure 2. Service tunnels are positioned above the main tunnels. They start from above the beam splitters and extend horizontally to the ET-HF input test masses or between the ET-LF and ET-HF end test masses, depending on the L or triangular configuration. These short tunnels serve a dual purpose: during construction, they provide access for drilling the small-diameter pilot boreholes used to excavate raise-bore wells, as shown in Figure 3; during operation, they house the upper stages of the ET-LF seismic isolation chains.

All seismic isolation chains are placed inside wells extending up just as much as needed by the specific chain. They are sufficiently wide to allow a generous tuning range along the optical axis. A vertical separation of several tens of meters between the service and main tunnels allows the implementation of significantly longer pendulum stages in the isolation chains. Longer pendulums shift the isolation roll-off to lower frequencies, thereby improving low-frequency noise performance and preserving margin for future upgrades. Because the pilot boreholes required for raise-boring are low-cost, even a large separation between the two tunnels is not expected to incur significant additional cost.

Locating the seismic isolation anchoring points well above the interferometer plane offers numerous technical and operational advantages, as discussed in the following sections. A related design was previously proposed by F. Amann et al.[10]. The scheme presented here extends this concept, with particular emphasis on ease of installation and maintenance, flexibility for future upgrades, and practical feasibility, while also providing significant potential benefits for interferometer operation and improved seismic activity monitoring.

All raise-bore wells—except those intended for personnel access—are blind, extending upward from the main tunnel toward the service tunnel but not connecting to it. Only the ET-LF test-mass pre-isolator, at the anchoring points of its seismic isolation chain, is located in the service tunnel, mounted directly on the rock floor for improved stability and reduced tilt sensitivity. Less demanding optics require shorter suspension systems. The upper stages of these suspensions are positioned on floors implemented near the top of the shorter wells. A compact inverted pendulum (IP), 1 m tall, is sufficient in all cases.

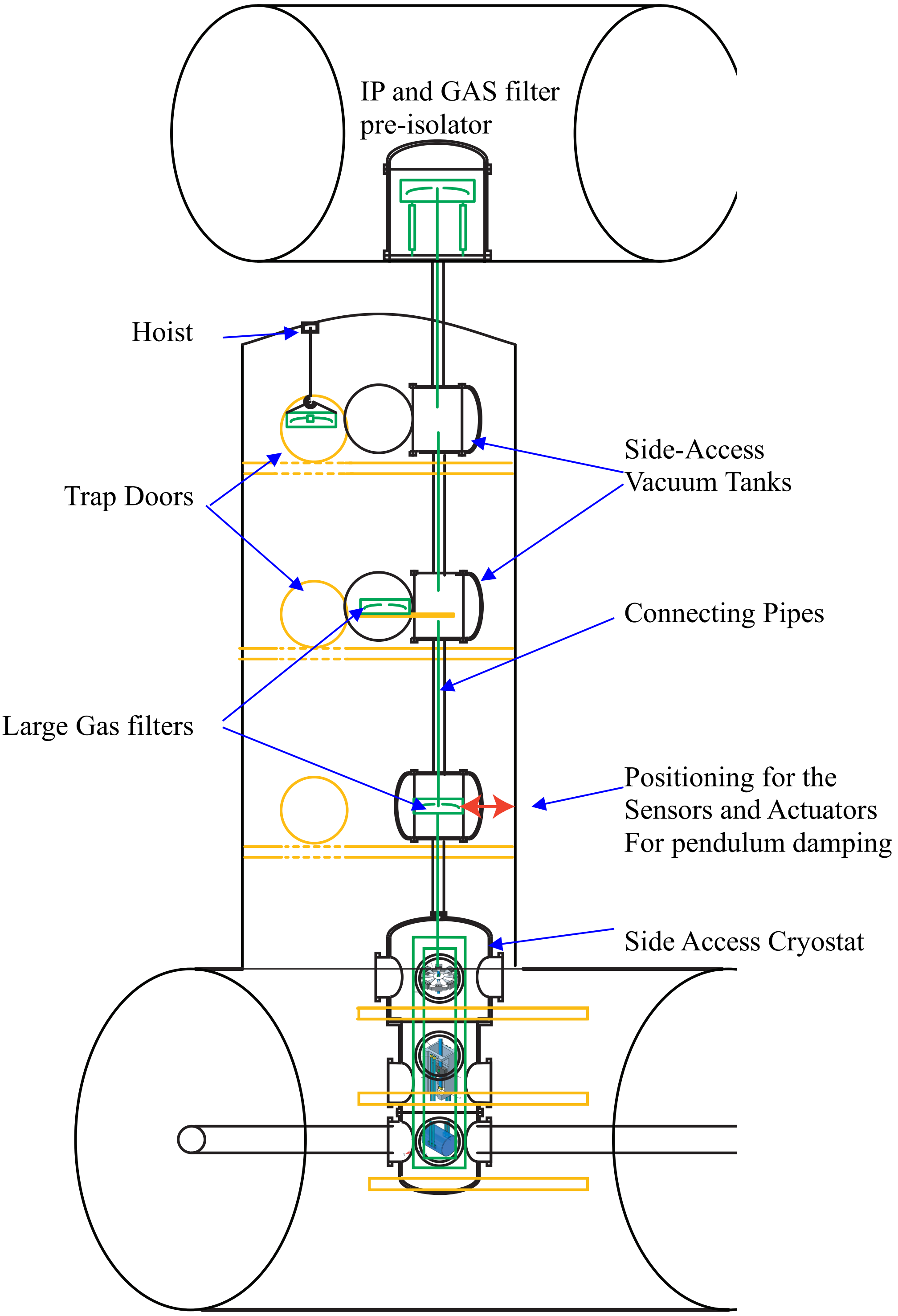

*Figure 2: Schematic illustration of the proposed system. The pre-isolator is mounted on the floor of the service tunnel. The seismic isolation filters are implemented in side-access vacuum chambers. The chambers are positioned on floors at appropriate levels within the raise-bore wells. The red arrows illustrate the position of one of the sensor actuator pairs referred to the rock wall that are used to damp the pendulum motion. The seismic isolation chain supports the cryogenic mirror suspensions located at the main-tunnel level*[11].

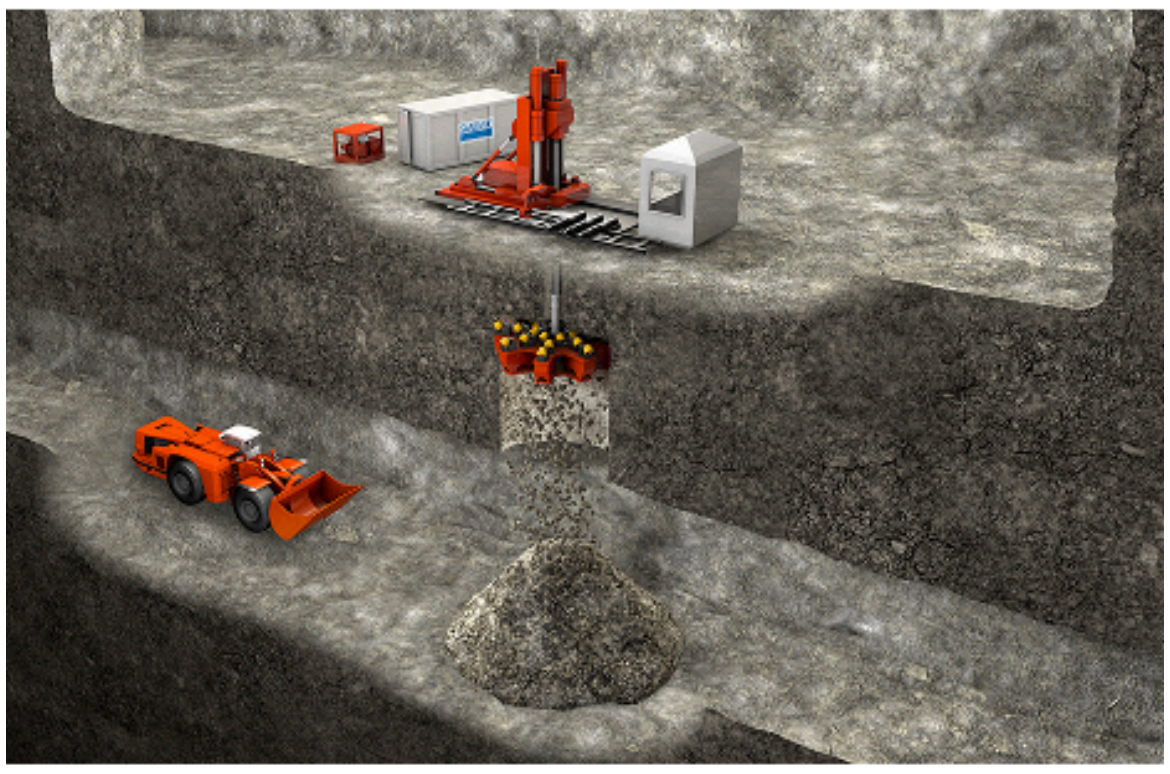 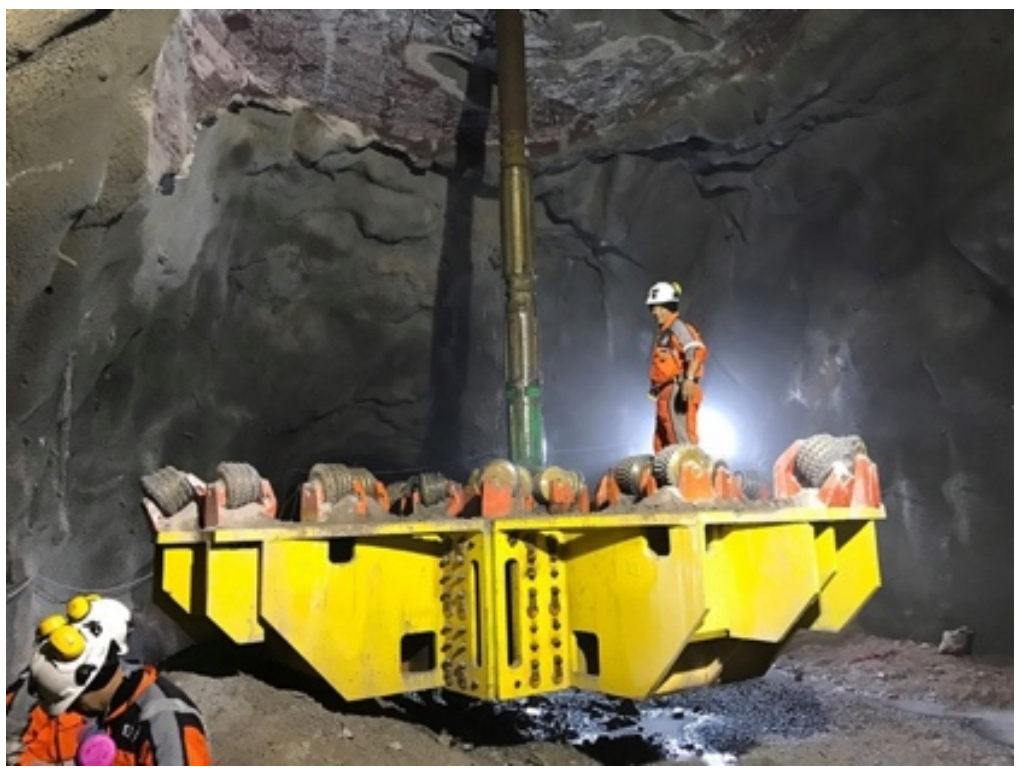

*Figure 3: Schematic of the raise-bore technique used to excavate wells of large diameter between two tunnels. A small-diameter pilot borehole is first drilled to introduce a drive shaft that supports the reaming shield. The shield is then rotated and pulled upward to enlarge the bore while falling debris is removed from below. By stopping the reaming before breakthrough, blind wells can be created. The same reaming head is reused for multiple wells. An actual raise-bore shield is shown on the right. The shield will be shaped to produce a dome-profile of the roof of the well.*

The pre-isolators are straightforward to install and service by lifting a bell-shaped vacuum chamber cap and working from the side. The mechanical and vacuum connection to the underlying suspension filters is through a small-diameter opening equipped with a pipe. Supporting all seismic isolation chains from above minimizes congestion in the space around the test mass and auxiliary optics. The improved lateral access, together with the reduced cryostat diameter and the cryogenic suspensions proposed in Ref. [11,13], enables taller test-mass suspensions to be assembled directly inside the cryostat, without the need for personnel to enter it, while allowing the work to be performed from a cleanroom environment within the tunnel. [12,13]

The shorter wells are also fitted with radially anchored floors positioned at appropriate heights to support vacuum chambers equipped with side-access flanges. These flanges provide safe and convenient access to the filters. The vacuum chambers are connected by small-diameter vacuum pipes.

Component handling within each well is simplified by a hoist capable of lifting vacuum chambers, filters, and other assemblies from the main tunnel to the designated level through trap doors. Once a trap door is closed, the components can be moved horizontally to their points of use. This arrangement eliminates the need for large bridge cranes. Personnel access to the different floor levels is provided by spiral staircases.

Noisy equipment such as cryogenic chillers can be located in adjacent tunnels, a few meters above and laterally offset from the main tunnel, allowing the heat links to be connected at the appropriate level while thick rock walls provide effective noise and vibration isolation.

## Rock-stable platforms

Real-time, high-precision measurements of rock motion using a new class of sensors enable correction signals to be applied to the inverted pendulums of all seismic isolation chains, including during seismic events[14].

The amplitude of seismic motion is generally smaller at depth, owing to the greater rigidity of the surrounding rock, the reduced contribution of Rayleigh waves, and the lower amplification associated with reflections from the surface. The combination of longer, lower-frequency pendulums and active compensation of seismic motion at the anchoring points of the isolation chains could enable the interferometer to withstand most earthquakes, with the detector temporarily suspending astrophysical observation but retaining optical lock to allow rapid recovery. Virgo has already shown the capability of maintaining lock this way during certain kinds of small earthquakes.

Maintaining lock during such events would not only improve the observational duty cycle and preserve the thermal stability of the mirrors but could also provide scientifically valuable geophysical data. The system would record both transient and permanent strains in the surrounding rock.

Accessing the experimental halls of surface detectors perturbs astrophysical observations and can cause loss of lock. As illustrated in Figure 4, physically separating the anchoring points of different seismic isolation chains into effectively independent caverns provides much greater isolation than is present between the control rooms and experimental halls of surface detectors. Because these are the only places where mechanical disturbances can be coupled into the suspended optics, maintenance or intervention on one interferometer would not be expected to disturb the astrophysical observing runs of the other(s). This advantage is, of course, not available in the large cavern concept where the support points of all interferometers are on the same floor in the same cavern and the tall structures amplify the perturbations.

## Optical levers

Aligning the two mirrors of a long Fabry-Perot optical cavity so that they face each other with sufficient precision to allow lock acquisition is a difficult task. In current interferometers it takes substantial time to reacquire lock, especially after long interruptions. Optical levers with beams reflecting from the test mass mirror surfaces have been implemented and then upgraded in LIGO to provide sufficient pointing memory to more rapidly realign the mirrors. With ET's 10- or 15-km-long arms, pointing with nano-radian precision will be necessary both in pitch and in yaw. This precision will most likely require that both the transmitting telescope and the position sensing detectors of the optical levers be directly mounted on rock, without extended support structures that would reduce the angular stability. Vacuum tubes extending from the cryostat to the rock wall will be necessary. Considering a main tunnel diameter of 8 m and an optical lever reflection angle of 15°, these pipes will be over 15 m long in the dual tunnel scheme. In the large cavern concept, the beams would have to be much longer, with the added constraint of having to clear the inverted pendulum support structures and local traffic.

The optical levers would also provide monitoring of the secular movement of the locked Fabry-Perot with respect to the rock.

## Advantages of separate vacuum chambers with side access

The side access vacuum chambers made possible by the tunnel-and-well configuration have several advantages.

The heavy ET test masses require vertical filters exceeding 1.5 m in diameter and weighing 200–250 kg. These large filters can only be safely installed and efficiently serviced in side-access vacuum chambers with suitable working floors.

Isolation filters for lighter, less-demanding optics are significantly smaller—comparable to the 70-cm KAGRA-type filters—and can be positioned in smaller vacuum chambers with at least ±2 m of longitudinal positioning freedom within the well diameter.

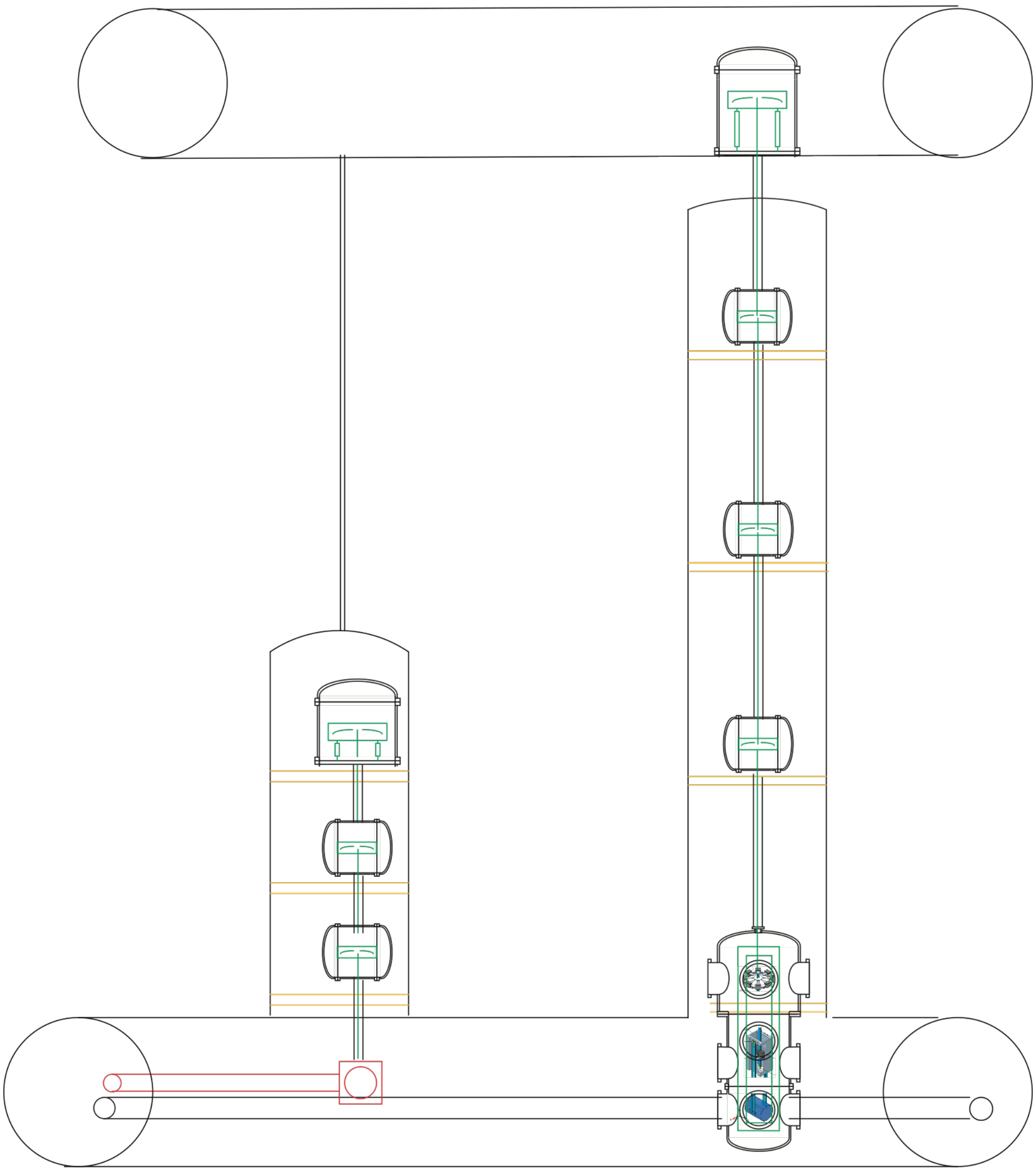

*Figure 4: Schematic illustration of the physical separation between suspension points and the main tunnel. With anchoring points relegated into separate alcoves in the rock, work on an interferometer, or in the main tunnel, can be expected to be possible without interfering with the other interferometer, even in observation mode.*

All vertical filters tuned to low frequencies exhibit strong temperature sensitivity, their individual vacuum chambers can be thermally controlled to maintain each filter at its optimal vertical operating point.

Virgo went to considerable effort to introduce a small vacuum conductance section between the seismic isolation chain and the mirror suspensions to allow differential pumping and reduce the strict vacuum compatibility requirements for the upper stages. The small-diameter pipes connecting the vacuum chambers can be designed with low vacuum conductance, easing differential pumping requirements for the upper stages. This configuration also mitigates the risk of accidental contamination of the cryostat.

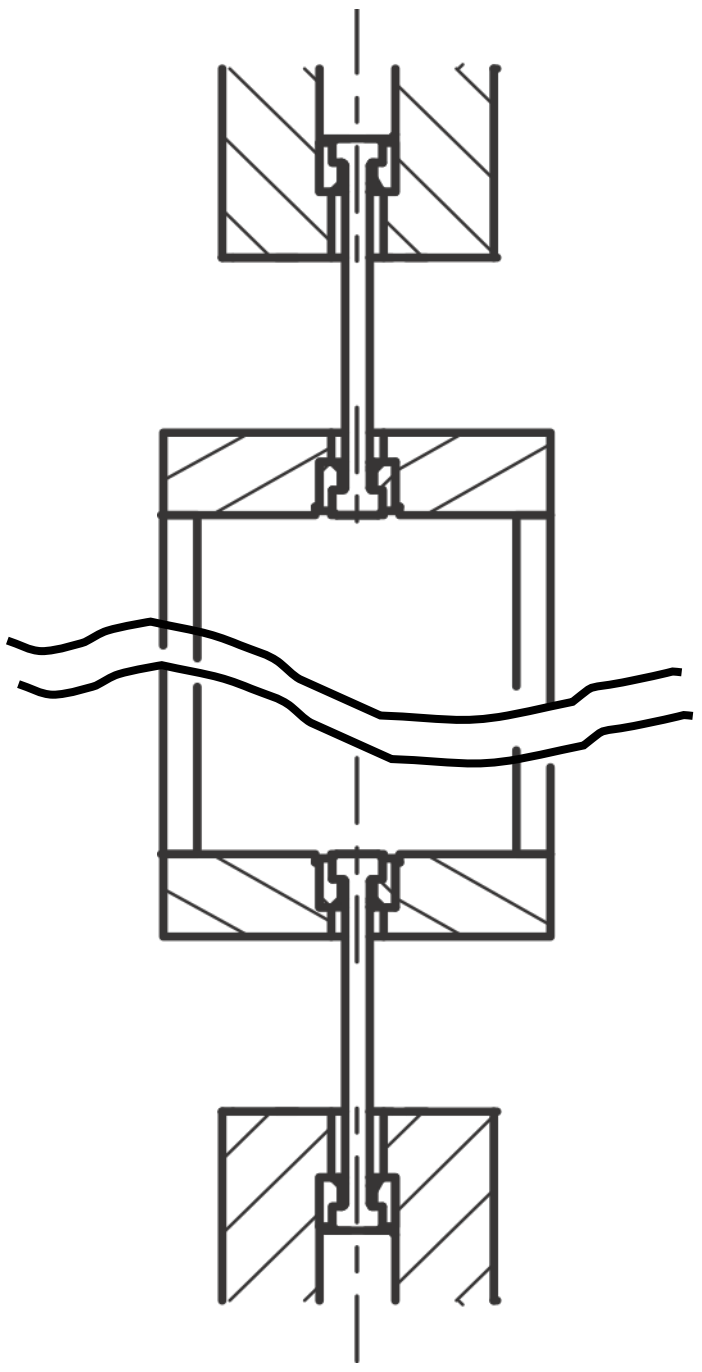

*Figure 5: Schematic illustration of the connection between filters. Two short double–nail-head wires connected by a rigid intermediate tube replacing a single long wire to achieve increased torsional stiffness while preserving pitch and roll softness.*

Very long suspension wires lead to undesirably low yaw resonant frequencies. In addition, fabricating long wires incorporating double-nail-head terminations is technically challenging and costly. To mitigate these issues in KAGRA, the suspension wires were machined to a thinner diameter only over suitable lengths near their attachment points. In ET, each long wire can be replaced by a pair of shorter wires attached to the ends of a long, lightweight tube that is rigid in torsion, as shown in Figure 5. This configuration preserves the effective pendulum length while increasing the yaw stiffness to an suitable level without compromising pitch, roll, or overall suspension performance. Implementation of these connections is within the side-access vacuum chambers

## Pendulum-Mode Damping

Although not a fundamental noise source, mirror control noise constitutes one of the most significant and persistent technical limitations to low-frequency gravitational-wave sensitivity. A substantial fraction of the actuator authority required during interferometer lock acquisition arises from the need to rapidly halt the test-mass pendulum motion within the narrow resonance width of the optical cavities (less than a nm). The 200 kg test mass may carry considerable momentum due to residual pendulum-mode velocity within the seismic isolation chain. The force $F_{lock}$ required to arrest this motion is given by

$$F_{lock} = \frac{2\mathcal{F}MV^2}{\lambda} \qquad (1)$$

where $\mathcal{F}$ is the interferometer finesse, $M$ is the 200 kg mirror mass, $\lambda$ is the laser wavelength, and $V$ the residual pendulum velocity. The mirror actuators must be sufficiently strong and fast to absorb the entire momentum of the test mass with a sub-millisecond impulse. In present interferometers, this step may take many attempts and considerable time to perform. Figure 6 shows the dependence of the required authority vs. oscillation amplitude using the equation 1, adapted for the ET-LF case[15].

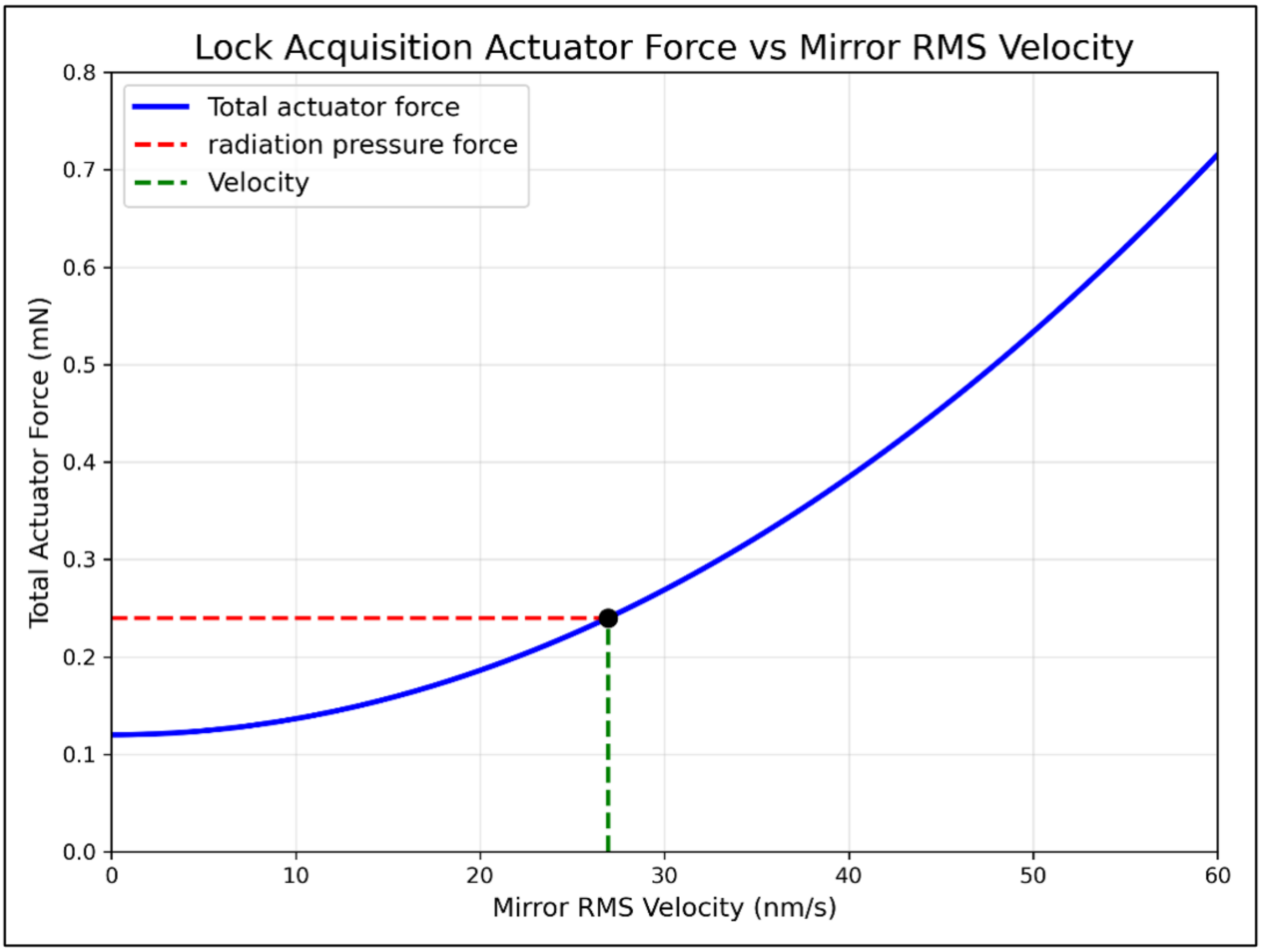

*Figure 6: Actuator authority required to achieve interferometer lock as a function of the test-mass pendulum velocity, according to equation 1. The baseline level at 0 nm/s is what required to balance the radiation pressure surge from the 18 kW of stored optical power in ET-LF. The green dashed line is the residual velocity that results in a doubled actuator authority requirement [16]. Damping to the Peterson low-noise model brings the residual velocity well below the green line. Of course in ET-HF the radiation pressure surge is much larger, and a gradual increase of optical power may be necessary in addition to the modal damping to reduce the required authority for lock acquisition.*

Actuators capable of delivering strong forces would, if operated during astrophysical observation, inevitably inject an excess of control noise. The actuator strength required to acquire interferometer lock is dramatically reduced if the residual velocity of the pendulum modes of the seismic isolation chain can be first reduced to sufficiently low amplitudes. This reduction can be achieved using sensor–actuator feedback systems operating between the filters of the seismic isolation chain and the rock wall. Modal damping is already applied in present interferometers, but to reduce the residual velocity to or below the Peterson low-noise model the optical length sensor used in the feedback must be connected directly to the bedrock. In this way a residual velocity of ~10 nm/s may be achieved, thus rendering the momentum negligible compared with the force impulse associated with the radiation-pressure surge at lock acquisition. Under these conditions lock acquisition would become substantially simpler, requiring minimal actuator authority and correspondingly a reduced control noise.

This technique cannot be implemented in large caverns, where in the most important longitudinal direction the rock is hidden by other vacuum tanks and/or too far away.

## Tilt, Translation and Strain Seismic Noise Sensors for Newtonian-Noise Subtraction

The filters of the seismic isolation chains are inertial masses that are orders of magnitude quieter than those employed in conventional seismometers. They can therefore be used to develop seismic activity sensors with sensitivities far exceeding those achievable with existing instruments, provided that the bedrock whose motion is to be measured is in close proximity.

Hosting the seismic isolation chains in wells makes it possible to employ short optical length sensors [17,18] to measure the bedrock motion, achieving sensitivities down to $10^{-14}$ m/√Hz in the ET sensitivity band, up to two orders of magnitude better than those of high-sensitivity seismometers. Furthermore, by using multiple filters within a single chain, it is possible to distinguish rock tilt from horizontal translational motion with high precision, measuring tilt at levels down to $10^{-14}$ rad/√Hz.

The differential signals from the filters supporting the optical elements provide a direct measurement of the strain in the rock on either side of the test masses to less than $10^{-15}$ 1/√Hz. Measuring the strain between test masses separated by 15 km at a level of $10^{-18}$ 1/√Hz and extending this measurement to zero frequency when the interferometer is locked, would be of significant interest for geophysical studies.

A preliminary discussion of these advanced seismic sensors is presented in ref. [12] while a detailed analysis will be the subject of a separate, forthcoming paper.

## Excavation Volumes

The excavation volumes required for the current L-shaped ET design are approximately 330,000 $m^3$ for the caverns alone [19]. In comparison, a 150 m-long service tunnel with a 6 m diameter requires less than 4,500 $m^3$ of excavation—or about 5,000 $m^3$ when the associated raise-bore wells are included. Even after accounting for all tunnel-and-well assemblies needed for the ET-LF and ET-HF beam splitters, the end mirrors, and the input, output, and vacuum squeezing optics, the total excavated volume is roughly an order of magnitude smaller than in the cavern option. This reduction is even greater for the triangular ET configuration, resulting in significantly shorter excavation times and lower overall excavation costs.

## Conclusions

We propose replacing the large ET experimental caverns with a more compact, dual-tunnel configuration in which raise-bored wells extend up from the main tunnel toward a short upper tunnel section. This design choice offers significant scientific, technical, and operational advantages.

The introduction of vertically separated tunnels enables the use of longer pendulums, lowering the isolation frequencies and providing additional margin against seismic and suspension thermal noise, thereby enhancing the scientific potential of the detector and providing robustness against future upgrades**.**

In addition, the mechanical stability of the surrounding rock substantially improves interferometer controls: can be achieved will be easier with reduced residual pendulum motion, and achieved using weaker actuators, which in turn minimizes control-noise injection. The stable rock and separation of the anchoring points of the seismic isolation chains into effectively independent rock alcoves will allow intervention on one interferometer with minimized disturbance of the other(s).

The dual-tunnel design also allows the separation of seismic motion from ground tilt and to measure strain around the test mass at unprecedented sensitivity for Newtonian-noise subtraction, a subject treated in a forthcoming article.

Operational benefits include a decluttered environment around the cryostats, improved side access for installation and maintenance of individual components through intermediate-floor workspaces, and greater flexibility for potential future modifications of the optical configuration within the wide-diameter wells.

Finally, the proposed configuration requires roughly an order of magnitude less excavation than even the reduced-size large-cavern alternative discussed in Ref. [9], offering substantial time and cost savings while improving the scientific performance.

## Acknowledgments

We thank Anna Greer, Antonio Perreca, Jan Harms, Paul Ophardt and Francesca Badaracco for useful discussions.

## Disclosures

The authors declare no conflicts of interest. No experimental data from external sources are used in this work. This work was performed for educational purposes and received no external funding.

## References

[1] Punturo et al., "The Einstein Telescope: a third-generation gravitational wave observatory," Class. Quantum Grav. 27 (2010) 194002.
M. Punturo, H. Lück, M. Beker, *"A Third Generation Gravitational Wave Observatory: The Einstein Telescope"* (in *Advanced Interferometers and the Search for Gravitational Waves*, Springer, 2014).

[2] Hild, S., Chelkowski, S., Freise, A., Franc, J., Morgado, N., Flaminio, R., De Salvo, R., “A Xylophone Configuration for a Third Generation Gravitational Wave Detector,” *Classical and Quantum Gravity*, Vol 27, No 1, 015003 (2010).
[3] Harms, Jan, et al. "Characterization of the seismic environment at the Sanford Underground Laboratory, South Dakota." *Classical and Quantum Gravity* 27.22 (2010): 225011.
F. Badaracco, J. Harms, “A lower limit for Newtonian-noise models of the Einstein Telescope,” *Eur. Phys. J. Plus* 137, 1250 (2022).
[4] González, Gabriela. "Suspensions thermal noise in the LIGO gravitational wave detector." *Classical and Quantum Gravity* 17.21 (2000): 4409.
[5] Levin, Y., “Internal thermal noise in the LIGO test masses: a direct approach”, *Physical Review D*, Vol. 57, Issue 2, pp. 659-663 (1998)
[6] Branchesi, Marica, et al. "Science with the Einstein Telescope: a comparison of different designs." *Journal of Cosmology and Astroparticle Physics* 2023.07 (2023): 068.
[7] E. Calloni et al., “*The Superattenuator: a mirror suspension system for reducing seismic noise in the VIRGO gravitational wave detector*,” *Nucl. Phys. B (Proc. Suppl.)* 54B, 310–314 (1997)
[8] Y. Aso, Y. Michimura, K. Somiya et al., *“Interferometer design of the KAGRA gravitational wave detector”*, *Phys. Rev. D* 88, 043007 (2013)
[9] Fiodor Sorrentino, et al., *The ET Baseline Detector Layout*, https://www.iris.unina.it/retrieve/f2d35064-6b7f-4029-a576-767d2216bafd/The_ET_Baseline_Detector_Layout.pdf , June 27, 2025
[10] Amann, Florian, et al. "Tunnel configurations and seismic isolation optimization in underground gravitational wave detectors." *Applied Sciences* 12.17 (2022): 8827.
[11] Arellano, Fabián E. Peña, et al. "A cryogenic test-mass suspension with flexures operating in compression for third-generation gravitational-wave detectors." *arXiv preprint arXiv:2503.19169* (2025). Accepted for publication in Classical and Quantum Gravity.
[12] Riccardo DeSalvo, et al. Dual tunnel configuration for the Einstein Telescope https://apps.et-gw.eu/tds/ql/?c=18627
[13] Harry Themann, et al., Animation of cryogenic suspension implementation in small diameter cryostat., https://apps.et-gw.eu/tds/ql/?c=18628
[14] Private communication, Paolo Ruggi, August 2026
[15] Diego Bersanetti. “Development of a New Lock Acquisition Strategy for the Arm Cavities of Advanced Virgo”. English. PhD Thesis. University of Genova, 2016. url: https://tds.virgo-gw.eu/ql/?c=13950.
[16] Nelson Leon, et el., Horizontal Seismic Motion and Damping Requirements for Improved Lock Acquisition of the ET-LF Cryogenic Suspension, https://apps.et-gw.eu/tds/ql/?c=18626

[17] Oliver Gerberding and Katharina-Sophie Isleif, Ghost Beam Suppression in Deep Frequency Modulation Interferometry for Compact On-Axis Optical Heads, https://www.mdpi.com/1424-8220/21/5/1708

[18] Tobias Eckhardt and Oliver Gerberding, Noise Limitations in Multi-Fringe Readout of Laser Interferometers and Resonators, https://www.mdpi.com/2673-8244/2/1/7

[19] Fiodor Sorrentino, et al., The ET Baseline Detector Layout by Design Task Force https://apps.et-gw.eu/tds/ql/?c=18214